\documentclass[preprint,showpacs,preprintnumbers,amsmath,amssymb,superscriptaddress]{revtex4}
\usepackage{amssymb}
\usepackage{amsmath}
\usepackage{graphicx}
\usepackage{txfonts}% Include figure files
\usepackage{subfigure}
\usepackage{dcolumn}% Align table columns on decimal point
\usepackage{bm}% bold math

\begin{document}b

%\begin{CJK*}{GBK}{song}
\title{Two-dimension plasma expansions with anisotropic pressure}% Force line breaks with \\
\author{Yongsheng Huang }
\email{huangyongs@gmail.com}
\author{Yuanjie Bi }
\affiliation{China Institute of Atomic Energy, Beijing 102413, China.}%

\affiliation{Department of Engineering Physics, Tsinghua University, Beijing 100084, China.}%
\author{Xiaojiao Duan }
\author{Naiyan Wang }
\author{Xiuzhang Tang}
\affiliation{China Institute of Atomic Energy, Beijing 102413,
China.}
\author{Zhe Gao}
\affiliation{Department of Engineering Physics, Tsinghua University, Beijing 100084, China.}%

%\author{Wang Naiyan$^2$}
%\author{Wang Naiyan$^1$}

%\author{Huang Yongsheng}
%\affiliation{China Institute of Atomic Engineering.

\date{\today}% It is always \today, today,
             % but any date may be explicitly specified

\begin{abstract}
A two-dimension self-similar solution is proposed for a plasma
expansion with anisotropic pressure. With the solution, it depends
on the relationship between the ratio of the longitudinal and the
transverse temperature of the plasma, $\kappa^2$ and the
electron-ion mass ratio, $\mu$, that the plasma front is composed by
a part of hyperbolic (or a plane) and a small pointed projection at
the center or a part of an ellipse. Zhang and coworkers's
experiments (PRL, 99, 167602 (2007))support our results for
$\kappa^2\in(\tau,1]$. For $\kappa^2\leq\tau$, there is an anomalous
high-energy plasma emission at the angle of near $90^{\text{o}}$ due
to longitudinal Coulomb explosion.
\end{abstract}

\maketitle

%\section{\label{sec:level1}Introduction}

The picosecond and femtosecond laser ablation\cite{ZhangNan},  the
ultrafast diagnostics of hydrodynamics \cite{Aliverdiev}, the
laser-ion acceleration\cite{Nature Schwoerer}, etc., attract more
and more international attention since the great development of
ultrashort and ultraintense laser technologies. The theory of plasma
expansions is not only the fundamental mechanism of laser ablation
but also the laser-ion acceleration. Although various theory such as
phase explosion \cite{Lorazo}, Coulomb explosion \cite{Roererdink},
self-similar plasma expansions \cite{P.Mora2,HuangAPL1},
two-dimension self-similar plasma expansions \cite{HuangAPL2}, and a
relativistic model\cite{M Passoni} have been proposed, the observed
angular-ion distributions such as ring structure
\cite{E.L.Clark2000}, time-resolved elliptic shadowgraphs of
material ejection \cite{ZhangNan} and coronal hydrodynamics of
laser-produced plasma\cite{Aliverdiev} have not been explained
analytically. The problems are that: the electrons generated in the
laser-solid interactions are anisotropic generally, however the
self-similar expansion \cite{P.Mora2,HuangAPL1} and the relativistic
model are both one-dimension, the two-dimension theory is only valid
for a plasma with isotropic pressure, therefore, they can not
predict the shapes of plasma front with anisotropic pressure.

\begin{figure}
{
\includegraphics[width=0.45\textwidth]{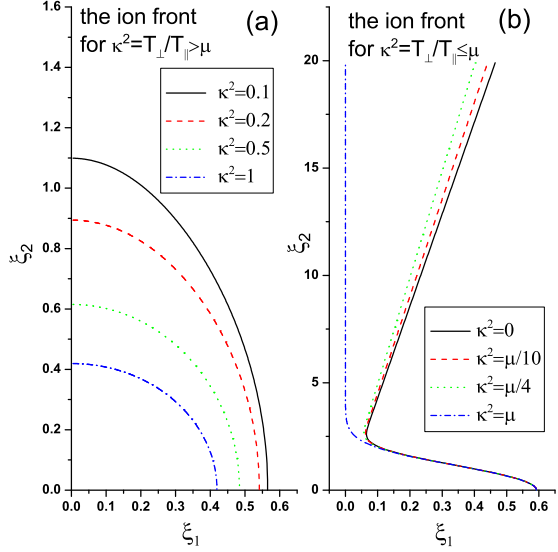}% Here is how to import EPS art
}  \caption{\label{fig:ionfront} (Color online) $(\xi_1,\xi_2)$ at
the ion front for different $\kappa^2$, the ratio of $T_{\bot}$ and
$T_{\parallel}$, which is compared with $\mu=Zm_e/m_i$, the mass
ratio of electron and ion for $\phi^0=1$ and $C_0=0$ in Eq.
(\ref{eq:ion front}). Figure \ref{fig:ionfront}(a) and (b) are
obtained from the solutions of Eq. (\ref{eq:ion front}). In Figure
\ref{fig:ionfront}(a), $\kappa^2=T_{\bot}/T_{\parallel}>\mu$. In
these cases, the curve of the ion front is a part of an ellipse,
especially, a cycle for $\kappa^2=1$. The major axis of the ellipse
is in the lower-temperature direction.  In Figure \ref{fig:ionfront}
(b), $\kappa^2=T_{\bot}/T_{\parallel}\leq\mu$. In this case, the ion
front contains two parts. For $\xi_2\geq2.5$, the ion front is a
part of a hyperbola or a plane ($\kappa^2=\mu$). For $\xi_2\leq2.5$,
the ion front is a small pointed projection. }
\end{figure}

In this letter, a self similar two-dimension solution for a plasma
expansion with anisotropic pressure is obtained. Although
isothermal-expansion assumption is used, the solution is still
suitable for a system whose temperature changes slowly, since the
whole process can be separated to several periods and the
temperature can be assumed as a constant in any period. In the
solution, there are two important parameters: the ratio of the
longitudinal temperature, $T_{\bot}$ and the transverse temperature,
$T_{\parallel}$ of the plasma, $\kappa^2=T_{\bot}/T_{\parallel}$,
and the electron-ion mass ratio, $\mu=Zm_e/m_i$, where $Z$ is the
charge number of the ion, $m_e (m_i)$ is the electron (ion) mass.
With the solution, it is found that the relationship between
$\kappa^2$ and $\mu$ decides the shape of the ion front: a
complicated surface or an ellipse shown by Figure
\ref{fig:ionfront}. For $\kappa^2\in[0,\mu)$, the ion front is
composed by a part of a hyperbolic and a small pointed projection at
the center. In the critical case, $\kappa^2=\mu$, the ion front is a
plane and a small pointed projection at the center. For
$\kappa^2\in(\mu,1]$, the ion front is a part of an ellipse and the
major axis is in the lower-temperature axis. The ion velocity at the
ion front has the similar construction to that of the ion front.
However, for $\kappa^2\in(0,1)$, the major axis of the ion-velocity
ellipse is in the higher-temperature axis. The difference of the
angular-energy distribution from the known
one\cite{R.A.Snavely2000,HuangAPL2} is that the energy is a delta
function at the maximum angle of near $90^{\text{o}}$ for
$\kappa^2\leq\mu$. It is an anomalous high-energy plasma emission at
the angle of near $90^{\text{o}}$ due to the longitudinal Coulomb
explosion. The plasma expansion is adiabatic and the temperature
changes slowly in a vacuum. However, there is energy transmission
from the plasma to the air besides that between the two directions
of the plasma. With the assumption: the plasma temperature decreases
with time exponentially in the air\cite{Wang}, the dependence of the
position of the ion front in the central axis on time is compared
with that observed by Zhang and coworkers's in \cite{ZhangNan}. From
the comparison, it is inferred that the plasma front shown in
\cite{ZhangNan} should be the front of $\text{Al}^{+}$,
$\text{Al}^{2+}$ or $\text{Al}^{3+}$.

%\section{\label{sec:level1}Time-Dependent Target Normal Sheath Acceleration}

We assume the pressure tensor of a two-dimension anisotropic plasma
satisfies :
\begin{equation}\label{pres0}
\overset{\rightharpoonup
\rightharpoonup }{P}=\left(
\begin{array}{cc}
n_ek_BT_{\parallel} & 0 \\ 0 & n_ek_BT_{\bot}
\end{array}\right).
\end{equation}
if the ion temperature $k_BT_i\ll k_BT_{\parallel}$ and $k_BT_i\ll
k_BT_{\bot}$, where $T_{\parallel}$ is the electron temperature in
the x direction and $T_{\bot}$ is the electron temperature in the y
direction which is perpendicular to x. In the following discussion,
we will show the importance of the ratio, $T_{\bot}/T_{\parallel}$,
and the dependence of the shape of the plasma front on the ratio. In
the calculation of the formula of the ion front, we assume the
plasma expansion is isothermal. However, if the temperature changes
with time slowly, in any interval short enough, the temperature can
also be considered as a constant and our calculation and results
still hold. Therefore, to a certain extent, our calculation and
results allow a wide latitude of generality in applications.

For convenience, the physical parameters: the time, $t$, the length
coordinate, $x(y)$, the ion (electron) velocity, $v_i (v_e)$, the
electron field, $E$, and the ion (electron) density, $n_i(n_e)$ are
normalized by the inverse of the equivalent plasma frequency,
$\omega_{pi0}=\sqrt{{n_{e0}e^2}/{m_i\epsilon_0}}$, the equivalent
plasma Debye length, $\lambda_{D0}=c_s/\omega_{pi0}$, the equivalent
ion acoustic speed, $c_s=\sqrt{{Zk_BT_{e}}/{m_i}}$,
$E_0={k_BT_{e}}/{e\lambda_{D0}}$, and the reference hot-electron
density, $n_{e0}$ respectively, where $m_i$ is the ion mass, $Z$ is
the charge number of the ion, $e$ is the elemental charge and
$T_{e}=T_{\parallel}+T_{\bot}$ is an equivalent temperature. Then
the electric potential is normalized as ${\phi}=e\psi/k_BT_{e}$,
where $\psi$ is the physical potential. $t, x (y), v_i (v_e), E, n_i
(n_e)$ are still used to represent the normalized parameters in the
following discussion.

%\subsection{\label{sec:level2}Isothermal Expansion}
In the new frame: $\tau={t}, \xi_1={x}/R_1(t), \xi_2=y/R_2(t)$ , the
ion (electron) velocity\cite{HuangAPL2} satisfies:
$v_{i,x}(v_{e,x})=\xi_1R_1^{'}, v_{i,y}(v_{e,y})=\xi_2R_2^{'}$. The
ion (electron) density is assumed as:
$n_{i(e)}=N_{i(e),1}(\xi_1,\xi_2)/R_1^2+N_{i(e),2}(\xi_1,\xi_2)/R_2^2
$, which is similar to that in \cite{HuangAPL2}. With the
transformation, the equations of continuity and motion, and
Poisson's equation are obtained easily in the new coordinate system.
The condition for the automatical satisfiability of the continuity
equation is $R_2/R_1=\kappa$, where $\kappa$ is a constant. Then the
ion (electron) density is simplified to
$N_{i(e)}(\xi_1,\xi_2)/R^2(t)$, where
$N_{i(e)}=N_{i(e),1}+N_{i(e),2}/\kappa^2$, $R=R_1$. Solving the new
motion equation of ions gives the potential in the ion region,
$\phi=-\phi^0(\xi_1^2+\kappa^2\xi_2^2)$, where $\phi^0$ is a
constant.  Therefore the electron density in the ion region
satisfies: $N_e(\xi_1,
\xi_2)=N_{e,0}\exp[-(1+\mu)\phi^0(\xi_1^2+\xi_2^2)/\alpha_1]$ by
solving the electron motion equation with
$\kappa^2=T_{\bot}/T_{\parallel}$, where
$\alpha_1=T_{\parallel}/T_{e}$. From Poisson's equation, the ion
density is $N_i=N_e+4$. The electron motion equation can be solved
in the ion region due to the special form of the electric potential:
$\phi=\phi_1(\xi_1)+\phi_2(\xi_2)$. However, beyond the ion front,
the potential and electron density are governed by the motion
equation of electrons and Poisson's equation together, since the
potential can not be separated to $\phi_1(\xi_1)+\phi_2(\xi_2)$ and
then the electron density can not be solved from the motion equation
solely. Combined the motion equation and Poisson's equation, a
two-order partial differential equation of the electron density can
be achieved. The first integral of it gives: $ (\frac{\partial
Y}{\partial\xi_1})^2+\kappa^2(\frac{\partial
Y}{\partial\xi_2})^2=\frac{2\exp(Y)-4(1+\kappa^2)\mu\phi^0Y}{\alpha_1}+C_0$,
where $C_0$ is the first integral constant and $Y=\ln(N_e)$. The
physical condition requires that Y is a $C^1$ function and
therefore, the curve equation of the ion front is:
\begin{equation}\label{eq:ion front}
\begin{array}{cc}
\frac{\xi_1^2}{A^2}+\frac{\xi_2^2}{B^2}=D\exp[-\frac{(\xi_1^2+\xi_2^2)}{2D\phi^0}]+C_0.\\
A^{-2}=1-\mu\kappa^2, B^{-2}=\kappa^2-\mu,
D=\frac{1}{2(1+\mu)(1+\kappa^2)\phi^{0,2}}.
\end{array}
\end{equation}
where $C_0$ is a constant and $\phi^{0,2}$ is the square of
$\phi^0$. With Eq. (\ref{eq:ion front}), the curve of the ion front
may be some part of a hyperbola, an ellipse or a cycle approximately
for different ratio of the temperatures ( $T_{\bot}$ and
$T_{\parallel}$), $\kappa^2$. A critical value of $\kappa^2$ is the
mass ratio of electron and ion, $\mu$ or the inverse of it.

Due to the assumed symmetrical form of the pressure, we just need to
consider the cases for $\kappa^2=T_{\bot}/T_{\parallel}\in[0,1]$.
Figure \ref{fig:ionfront} shows the eight cases of the ion front for
$\kappa^2\in[0,1]$, which are obtained by solving Eq. (\ref{eq:ion
front}) with $C_0=0$ and $\phi^0=1$. For $\kappa^2>\mu$, the curve
of the ion front is a part of an ellipse whose eccentricity is
$e=\frac{(1-\kappa^2)(1+\mu)}{1-\kappa^2\mu}$ as shown by Fig.
\ref{fig:ionfront}(a). As $\kappa^2\rightarrow \mu$,
$\mathrm{e}\rightarrow 1$. Specially, for $\kappa^2=1$, i.e.,
$T_{\bot}=T_{\parallel}$, the ion front is a part of a cycle. For
$\kappa^2\leq\mu$, the ion front is complicated and contains two
parts: a part of a hyperbola and another curve as shown by Fig.
\ref{fig:ionfront}(b). For the part of the hyperbola, $\xi_2\geq2.5$
and the slope of the asymptote is
$\sqrt{\frac{1-\mu\kappa^2}{\mu-\kappa^2}}$. As
$\kappa^2\rightarrow\mu$, the slope trends to infinite and the ion
front becomes a flat surface as shown by the dash dot line in Fig.
\ref{fig:ionfront}(b).

Now imaging the following physical picture: the initial state is
$T_{\bot,0}=\mu T_{\parallel,0}$, and the temperature $T_{\bot}$
increases slowly to the equilibrium temperature, $T_{eq}$ ( Note:
$T_{eq}=(\mu+1)T_{\parallel,0}/2$ in our two dimensional case.
However, in the three dimensional case, if we assume the temperature
in the y direction is equal to that in the z direction at the very
beginning time, $T_{eq}=(2\mu+1)T_{\parallel,0}/3$. ), when the
thermodynamic equilibrium is reached, since the energy is
transmitted between the transverse direction x and the longitudinal
directions (Note: in two dimensional case, y or z is the
longitudinal direction; in the three dimensional case, y and z are
all longitudinal directions where we assume the temperature in the y
direction is equal to that in the z direction.) with collisions. At
the very beginning time, the ion front is a flat surface with a
small pointed projection at the center shown by the dash dot line in
Fig. \ref{fig:ionfront}(b). After an interval, $\kappa^2>\mu$, and
then the ion front develops to a part of ellipses shown by the
solid, dash, and dot line step by step in Fig.
\ref{fig:ionfront}(a), which are also shown by Figure 2 (b-h) in
\cite{ZhangNan}. At last, the ion front trends to a part of a cycle
shown by the dash dot line in Fig. \ref{fig:ionfront}(a), which is
also shown by outer layer of Figure 2(i) in \cite{ZhangNan}. That is
a depiction of the two dimensional plasma expansion into a vacuum.
However, it has not been verified experimentally in a vacuum. In
spite of that, the time-resolved shadowgraphs of material ejection
shown by Zhang and coworkers in \cite{ZhangNan} support our theory
without doubt. The outermost layer of the shadowgraph does undergo
the physical processes depicted by our above discussion. It is the
most difference between the experiment and our scene that the plasma
expansion is in the air or in a vacuum. In the air, there are
thermal transmission from the plasma to the air, collisions between
the plasma and the air molecules and other nonlinear effects, and
then the outermost layer of the plasma cools down. In this case, the
shock wave front (shown by Fig. 3 in \cite{ZhangNan}) forms since
the particles inner the plasma catch up with the particles at the
front. In a vacuum, these phenomenons can not happen.

\begin{figure}
{
\includegraphics[width=0.45\textwidth]{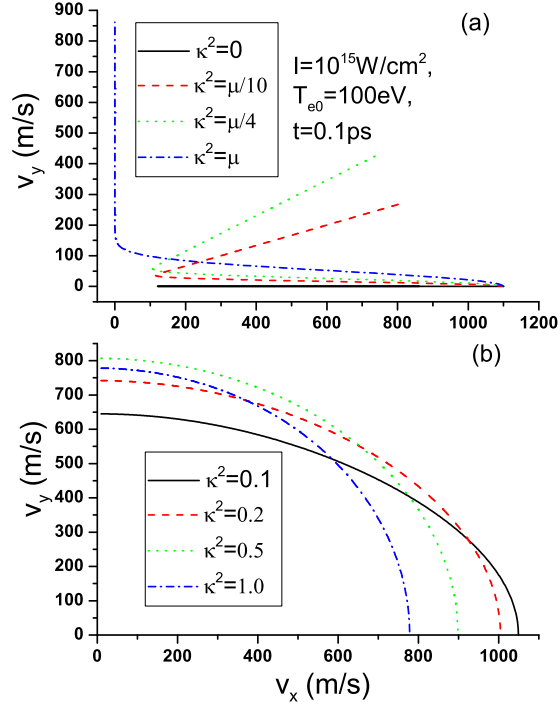}% Here is how to import EPS art
}  \caption{\label{fig:vdistri} (Color online) The ion velocity
distribution at the ion front for $I=10^{15}\mathrm{W/cm^2}$,
$\lambda=0.8\mathrm{nm}$, and
$\kappa^2=T_{\bot}/T_{\parallel}\in[0,1]$, $\phi^0=1$ in Eq.
(\ref{eq:ion front}), $\alpha(t)=0.003$ at $t=0.1\mathrm{ps}$. (a)
The ion velocity distribution at the ion front for
$\kappa^2\leq\mu$. (b) The ion velocity distribution at the ion
front for $\kappa^2>\mu$. }
\end{figure}
With the ion motion equation, $R(t)$ satisfies:
$\int^{\alpha}_0\exp(\alpha_1^2)d\alpha_1=\sqrt{\phi^0}t/R_0$, where
$\alpha=\sqrt{\ln(R/R_0)}$ and $R_0=R(t=0)$. Figure
\ref{fig:vdistri} shows the ion-velocity distribution at the ion
front obtained from our two-dimension theory for different
temperature ratio, $\kappa^2=T_{\bot}/T_{\parallel}\in[0,1]$. For
different times, the shapes of the ion-velocity distribution are the
same although the amplitudes are different. For
$\kappa^2\in(0,\mu)$, the velocity distribution has two parts:
ellipse-like at small $v_2$ and hyperbolic-like at large $v_2$,
which is shown by the dot line or the dash line in Fig.
\ref{fig:vdistri}(a). The solid line in Fig. \ref{fig:vdistri}(a)
shows all the ions move towards the x direction since the
temperature in the y direction is zero. The dash dot line in Fig.
\ref{fig:vdistri}(a) shows the critical case: $\kappa^2=\mu$, in
which the velocity in the x direction at large $v_2$ is zero and all
the ions move in the y direction. For $\kappa^2\in(\mu,1]$, the ion
velocity distribution is a part of an ellipse whose eccentricity is
$\sqrt{\mu(\kappa^{-2}-\kappa^{2})/(1-\mu\kappa^{-2})}$ and the
major axis is in the high temperature direction, x. At $\kappa^2=1$,
the ion-velocity distribution is a part of a cycle and isotropic as
shown by the dash dot line in Fig. \ref{fig:vdistri}(b).

\begin{figure}
{
\includegraphics[width=0.45\textwidth]{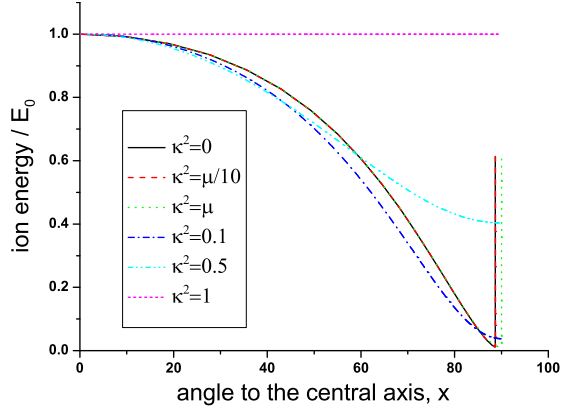}% Here is how to import EPS art
}  \caption{\label{fig:Edistri} (Color online) The ion energy
distribution at the ion front VS the angle to the central axis, x,
for $\kappa^2=T_{\bot}/T_{\parallel}\in[0,1]$, $\phi^0=1$ in Eq.
(\ref{eq:ion front}). In this figure, the energy for any angle is
normalized by $E_0$ for each $\kappa^2$, where $E_0$ is the ion
energy at the central axis.  }
\end{figure}

Figure \ref{fig:Edistri} shows the angular-energy distribution for
$\kappa^2\in[0,1]$. With it, the facts are indicated : (1) for
$\kappa^2\in[0,\mu]$, the energy decreases with angle except for the
maximum angle of near $90^{\text{o}}$. At the maximum angle, the
energy is a delta function with the peak value of about $0.62E_0$.
Especially, the maximum angle is $90^{\text{o}}$ for the critical
point, $\kappa^2=\mu$. This anomalous plasma emission happens since
the large Coulomb space-charge field dominates in the longitudinal
direction and is counteracted partly by the gradient of the pressure
in the transverse direction. (2) for $\kappa^2\in(\mu,1)$, the ion
energy decreases with angle monotonously. At $\kappa^2=1$, the
energy is a constant function with respect to angle.

\begin{figure}
{
\includegraphics[width=0.5\textwidth]{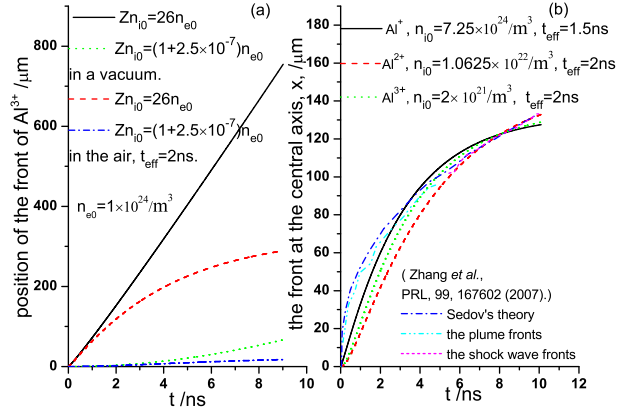}% Here is how to import EPS art
}  \caption{\label{fig:compp} (Color online) (a) The ion front of
$\mathrm{Al}^{3+}$ in the central axis,x, with $\xi_{1,f}=0.42,
\xi_{2,f}=0$, VS $t$, for $n_{e0}=10^{24}/\mathrm{m^3}$, and
$Zn_{i0}=26n_{e0}$, shown by the solid line (in a vacuum) and the
dash line (in the air, with the scale time of temperature,
$t_{eff}=2ns$), and $Zn_{i0}=(1+2.5\times10^{-7})n_{e0}$, shown by
the dot line (in a vacuum) and the dash dot line (in the air, with
the scale time of temperature, $t_{eff}=2ns$); (b) the front of
different ions in the central axis ,x, with
$\xi_{1,f}=0.42,\xi_{2,f}=0$, VS t. The solid line, the dash line
and the dot line correspond to the expansion in the air and are
obtained by our solution. The solid line is for $\mathrm{Al}^{+}$,
and $Zn_{i0}=7.25n_{e0}$, $t_{eff}=1.5\mathrm{ns}$. The dash line is
for $\mathrm{Al}^{2+}$, and $Zn_{i0}=1.0625n_{e0}$,
$t_{eff}=2\mathrm{ns}$. The dot line is for $\mathrm{Al}^{3+}$, and
$Zn_{i0}=2n_{e0}$, $t_{eff}=2\mathrm{ns}$. The dash dot line , the
dash dot dot line, the short dash line correspond to Figure 4 in
\cite{ZhangNan}.  }
\end{figure}

As discussed above, there is energy transmission from the plasma to
the air for plasma expansions in the air. Therefore, besides the
energy transmission between the longitudinal direction and the
transverse direction of the plasma, the temperature of the plasma
decreases exponentially to zero for the time long enough since the
temperature of the air is about $0\mathrm{eV}$. As an important
application, the position of the ion front in the air can be given
by our solution with the assumption: the temperature, $T_{e}$
decreases exponentially to zero\cite{Wang} with the scale time of
the temperature, $t_{eff}=-\partial \ln(T_{e})/\partial t$.
Therefore $T_{e}=T_{e0}\exp(-t/t_{eff})$ in the air and
$T_{e}=T_{e0}$ (which is $(T_{\bot0}+T_{\parallel0})$ in the two
dimensional case and $(2T_{\bot0}+T_{\parallel0})$ in the three
dimensional case where the longitudinal temperatures are the same.)
in a vacuum. The ions at the outermost layer are decelerated for the
plasma expansion in the air after a period of time shown by the dash
line and the dash dot line in Figure \ref{fig:compp}(a), however,
the velocity of them tends to a constant in a vacuum shown by the
dash line and the dot line in Figure \ref{fig:compp}(a).

Figure \ref{fig:compp}(a) shows different ion fronts in a vacuum, in
the air and for different initial charge-separation densities. The
difference of the ion fronts in a vacuum and in the air has been
discussed in the above paragraph. In the air, the collisions between
the plasma and the air molecules are important for the expansion and
considered gracefully through the exponential decrease of the plasma
temperature. The collisions result in nonlinear phenomenons and
energy transmission. For different initial charge-separation
densities of the plasma, the stronger the initial charge-separation
field and the more efficient the ions are accelerated as shown by
the comparison between the solid line and the dot line or between
the dash line and the dash dot line in Fig. \ref{fig:compp}(a). When
an ultra-short and ultra-intense laser pulse interacts with a solid
target, many interesting phenomenons: hot-electrons generation and
acceleration, x-ray and $\gamma$-ray emission, fast ions generation,
plasma ablation and plasma splash and so on occur. For different
phenomenon, the scale time is different. A fact is: the smaller the
scale time, the stronger the charge-separation field and the
nonlinearity in parameters of the plasma. For the generation of fast
ions, the scale time is about $1-10\mathrm{ps}$. However, for the
laser-plasma ablation, the scale time is several nanoseconds or even
longer. In the period of several pulse durations, the
charge-separation field is strong, i.e., the density difference,
$Zn_{i}-n_{e}$ (which is $4/R^2$ here.) is large (since $R(t)$
increases with time and tends to infinite as
$t\longrightarrow+\infty$.). Therefore, with the self-similar
solution for energetic ions generation given by Huang and coworkers'
in \cite{HuangAPL1}, the high charge-mass ratio, $Z/M$, ions are
accelerated preferentially and efficiently. When an ultra-intense
laser pulse interacts with a solid target,
$\text{H}^{+},\text{C}^{4+}, \text{C}^{3+}, \text{O}^{4+}$ and so on
are accelerated orderly. For Al target, $\text{Al}^{3+},
\text{Al}^{2+}, \text{Al}^{+}$ are accelerated lastly. According to
our analysis, the front surface observed by Zhang and coworkers' in
\cite{ZhangNan} should be the ions of $\text{Al}$ and the density of
the ions should be low since the scale time is of nanoseconds. The
fronts of the ions: $\text{Al}^{3+}, \text{Al}^{2+}, \text{Al}^{+}$
in the central axis, x, with $\xi_{1,f}=0.42, \xi_{2,f}=0$ are
calculated and compared with the curves given by Zhang and
coworkers' in \cite{ZhangNan} as shown by Figure \ref{fig:compp}
(b). They are consistent.

%\section{\label{sec:level1}Conclusion}
In conclusion, the analytic solution for the ion front of plasma
expansion with anisotropic pressure is predicted. It is supported by
experiments\cite{ZhangNan,Aliverdiev} lately. The ion-velocity
distribution at the ion front, angular-energy distribution at the
ion front are both given. It needs to be concerned that the energy
is a delta function at the maximum angle of near $90^{\text{o}}$ and
the plasma emits anomalously due to the longitudinal coulomb
explosion for $\kappa^2\in[0,\mu]$ as shown by the solid, dash, dot
lines in Fig. \ref{fig:Edistri}. In a vacuum, the expansion is
adiabatic, however, in the air, the temperature decreases with time
exponentially and collisions happen, and then the ions at the
outermost layer are decelerated and shock waves generate in the
front of the plasma front. From the comparison (shown by Fig.
\ref{fig:compp}) , it is concluded that the ions observed in
\cite{ZhangNan} are the ions of Al. However, the collisions and the
nonsymmetric components of the pressure tensor of plasmas have not
been considered directly in the calculation. Therefore, how the
energy transmission between the longitudinal direction and the
transverse direction or between the plasma and the air can not be
described in detail. In the air, they are challenges for our future
work.

This work was supported by the Key Project of Chinese National
Programs for Fundamental Research (973 Program) under contract No.
$2006CB806004$ and
 the Chinese National Natural Science Foundation under contract No.
$10334110$.

%\newpage %Just because of unusual number of tables stacked at end

%\end{CJK*}

\end{document}